\documentstyle[12pt]{article}
\input epsf
\newcommand{\be}{\begin{equation}}
\newcommand{\ee}{\end{equation}}
\newcommand{\ba}{\begin{eqnarray}}
\newcommand{\ea}{\end{eqnarray}}
\newcommand{\baa}{\begin{eqnarray*}}
\newcommand{\eaa}{\end{eqnarray*}}
\newcommand{\bb}{\begin{thebibliography}}
\newcommand{\eb}{\end{thebibliography}}

\newcommand{\bi}[1]{\bibitem{#1}}
\begin {document}
\begin{titlepage}
\begin{center}
{\Large \bf Impact-picture predictions \\[0.5cm]
       for the $\gamma\gamma$ total cross section at LEP}\\[1cm]
{C. Bourrely, J. Soffer}\\[0.3cm]
{\it Centre de Physique Th\'eorique \footnote{Unit\'e Propre de Recherche
7061}\\
CNRS, Luminy Case 907, F-13288 Marseille Cedex 09 - France},\\[0.5cm]

{Tai Tsun Wu}\\[0.3cm]
{\it CERN - Geneva, Switzerland and Gordon McKay Laboratory,\\
                     Harvard University, Cambridge, MA02138 - USA }
\end{center}
\vskip 1.0cm
\begin{abstract}
We show that the rising total cross section 
$\sigma(\gamma\gamma \rightarrow hadrons)$ 
recently observed by the L3 and OPAL Collaborations at LEP are fully 
consistent with the impact-picture for high-energy scattering.
The impact picture is then used to predict this total cross section at
higher energies.
These experimental results confirm once more the success of the theoretical
approach, which predicted for the first time, nearly thirty years ago, 
the universal increase of total cross sections at high energies.
\end{abstract}

\vskip 7cm


\noindent Number of figures : 1

\smallskip

\noindent March 1999

\noindent CPT-99/P.3793

\noindent CERN-TH/99-76

\noindent anonymous ftp: ftp.cpt.univ-mrs.fr\\
\noindent Web address: www.cpt.univ-mrs.fr
\end{titlepage}

\newpage

\smallskip

\begin{figure}
\epsfxsize=15cm
\centerline{\epsfbox{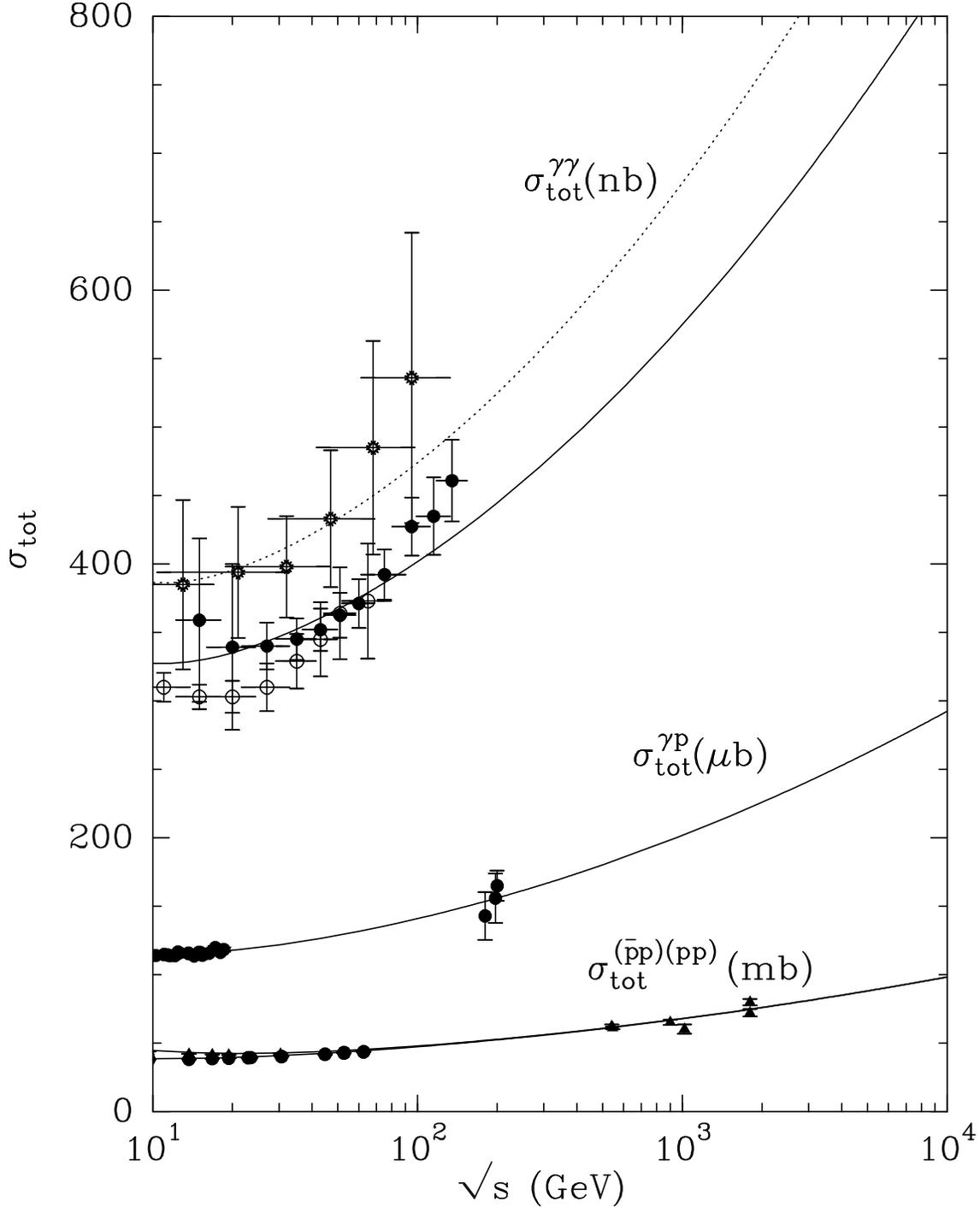}}
\caption{$\gamma \gamma$, $\gamma p$ and $\bar pp (pp)$ total cross sections as
a function of the center of mass energy $\sqrt s$, which stands respectively
for $W_{\gamma \gamma}$, $\sqrt s_{\gamma p}$ and $\sqrt s_{pp}$. Note that we
have used three different units. The bottom curves were calculated in
Ref.\cite{BSW84}
and for the data points (close circles for $pp$ and triangles for $\bar p p$)
see Ref.\cite{Pruss}. The middle curve was calculated in Ref.\cite{BSW94} and
the higher energy data points are from Refs.\cite {H1,ZEUS}. The top curves are
the impact-picture prediction compared to the LEP data, Ref.\cite{L3a} (open
circles), Ref.\cite {L3b}(close circles, preliminary data), solid curve
with $A_L = 8.5 . 10^{-6}$ and Ref. \cite{OPAL} (stars, preliminary data) 
dotted curve with $A_O = 10^{-5}$.}
\label{fi:sigtot}
\end{figure}

Recently at LEP the L3 and OPAL Collaborations have produced new results on the
total cross section $\sigma(e^+e^- \rightarrow e^+e^- hadrons)$
for several values of the $e^+e^-$ center of mass energy, up to 
$\sqrt s = 183 GeV$ [1-3].
The analysis of these data allows to isolate the two-photon 
cross section $\sigma(\gamma\gamma \rightarrow hadrons)$, which has been 
obtained in the range $5\leq W_{\gamma\gamma}\leq 145GeV$. 
The lowest energy data \cite{L3a}, which corresponds to 
$5\leq W_{\gamma\gamma}\leq 75GeV$, has revealed experimentally for the first
time a rise for the two-photon cross section. 
This increasing energy dependence has been confirmed on the full energy range,
as seen on Fig.~\ref{fi:sigtot}, showing that the photon behaves pretty 
much like a hadron.

Increasing total cross sections were first predicted nearly 
thirty years ago \cite{CW} entirely on theoretical grounds based on 
quantum field theory.  Indeed, one of the starting point of that 
theoretical work was $\gamma\gamma$ scattering; see for example 
\cite{CW69} and \cite{CW71}.  One of the results of this theory, 
sometimes referred to as the impact picture, is that there is a universal 
increase of all total cross sections at very high energies \cite{CW, CWb};
see Eq.(\ref{ssym}) below. It is the purpose of this letter to return 
to the root of the theory and apply it \cite{CWW, BSW79} for comparison 
with the experimental data on $\gamma\gamma$ total cross section
$\sigma^{\gamma\gamma}_{tot}$.

In view of the recent experimental results [1-3], several theoretical attempts have been made to explain the behavior of $\sigma^{\gamma\gamma}_{tot}$.
This energy behavior can be described by a Regge-type parametrization based 
on the exchange of Regge trajectories and Pomeron in the $t$-channel,
which was used for all hadron total cross sections \cite{DL}.
In early impact-picture predictions \cite{CWW}, a simple $s$ dependence
$s^{0.08}$ was first obtained and later extensively used by several authors
{\it i.e.} Ref.\cite{DL, PDG}. However, a best fit to the LEP data
by the L3 collaboration gives a higher power value \cite{L3b}. 
The Dual Parton Model with unitarization
constraint \cite{Eng} gives a faster rise of the cross section, also the model
of $\gamma \gamma$ scattering \cite{Schu} where the total cross section 
receives contributions from three event classes, VDM processes, direct and
anomalous processes.
In the framework of eikonalized amplitudes, a model
of mini-jet \cite{Corset} explains the increase of the cross section through
the rise of jet cross sections,  while a model \cite{Block} using the vector 
dominance and an eikonalized form of the quarks and gluons interactions, reproduces the energy dependence of the total cross sections for 
$pp$, $\bar pp$, $\gamma p$ and $\gamma \gamma$.

Returning to the impact picture, which is the basis of the 
present consideration, we have learned that the effective interaction 
strength increases with energy in the form \cite{CW, CWb}
\be
{s^{1+c} \over (\ln s)^{c'}}~,
\label{ssym}
\ee
a simple expression in terms of two key parameters $c$ and $c'$.
It should be emphasized that these two parameters are independent of 
the scattering process under consideration, i.e., the increase in 
the total cross section for example is universal.  The scattering 
amplitude from the impact picture is given by
\be
a^N(s,t) = is \int_{0}^{\infty} J_0(b\sqrt{-t})(1- e^{-\Omega(s,b)})bdb~,
\label{ampli}
\ee
where
\be
\Omega(s,b) = S_0(s)F(b^2) + \Omega_R(s,b)~,
\ee
$\Omega_R(s,b)$ being a Regge background which allows to use the model 
at rather low energy. 
The energy dependence is given by the crossing symmetric 
version of Eq.(\ref{ssym}),
\be
S_0(s) = {s^c \over (\ln s)^{c'}} + {u^c \over (\ln u)^{c'}}~,
\ee
$s$ and $u$ are the Mandelstam variable. The $t$-dependence of $a^N(s,t)$
is controlled by $F(b^2)$ whose Fourier transform is taken to be
\be
\tilde F(t) = f[G(t)]^2[(a^2 + t)/(a^2 -t)]~,
\ee
where $G(t)$ is given by
\be
G(t) = {1 \over (1 -t/m_1^2)(1 -t/m_2^2)}~.
\ee

It describes successfully $\bar pp$ and $pp$ elastic
scattering up to ISR energies, including the total and differential cross
sections, the polarization, and the forward real part of the amplitude, and
a systematic study of the experimental data available up to 1979 led to the values,
\be
c=0.167,~~~~~~~~~~~~~~c' = 0.748.
\label{cc}
\ee
Its predictions at very high energy, are in excellent
agreement with the data from the CERN SPS collider and the FNAL Tevatron, as we
recall for total cross sections in the bottom part of Fig.~\ref{fi:sigtot}, and
some others, at several $TeV$, which remain to be checked at the Large Hadron
Collider under construction at CERN. 

For hadron-hadron processes at high energies, the physical picture is such 
that each hadron appears as a black disk with a gray fringe, where the 
black disk radius increases as $lns$.
So far we have encountered the expanding proton in $\bar pp$ and $pp$ 
scattering, but also recently in a very different experimental situation, 
namely in $\gamma p$ scattering at HERA \cite{BSW94}. 
It can be shown that the energy rise observed in the $\gamma p$
total cross section, for the center of mass energy up to  $\sqrt s = 180 GeV$,
is entirely consistent with the theory of expanding protons \cite{CWb} and this
is also shown in the middle part of Fig.~\ref{fi:sigtot}.

 Similarly, since the parameters $c$ and $c'$, which control the 
increase of the total cross sections, are given by Eq.(\ref{cc}) and are 
universal for all scattering processes, we expect the high-energy 
behavior of $\sigma^{\gamma\gamma}_{tot}$ to follow approximately what we 
have obtained for $\sigma^{pp}_{tot}$, the $pp$ total cross section.  Accordingly, a simple way to obtain $\sigma^{\gamma\gamma}_{tot}$ is to use
the following approximate relationship
\be
\sigma^{\gamma\gamma}_{tot}(W_{\gamma \gamma}) = A \sigma^{pp}_{tot}(\sqrt s) ~,
\ee
where A is a normalization constant and $\sqrt s $ is the $pp$ center of mass
energy. Since there seems to be a normalization discrepancy between L3
and OPAL, the accurate determination of this constant is not needed for
comparison with experiments.
We have found that to get the best agreement with L3 data, 
$A_L = 8.5 . 10^{-6}$ is required and the use of 
Eq. (\ref{ampli}) (see also \cite{BSW84}) yields the
$\sigma^{\gamma\gamma}_{tot}$ shown as the solid curve in the top part of 
Fig.~\ref{fi:sigtot}.
Another possible choice which agrees with the
OPAL data is $A_O = 10^{-5}$, the corresponding dotted curve is also
shown in Fig.~\ref{fi:sigtot}.
We also display in Fig.~\ref{fi:sigtot} our predictions over a much higher 
energy range.  These prediction may be checked at a future $e^+ e^-$ 
linear collider. The success of our previous predictions for the $pp$
total cross section gives confidence for the present one for 
the $\gamma \gamma$ total cross section.

Therefore this universal energy rise, presented in Fig.~\ref{fi:sigtot}, 
for three different reactions, is one of the properties of the impact-picture 
approach which is once again verified by experiment. 
We expect that more accurate data and possible access to higher energy 
domains will strenghten the validity of these predictions.\\ \\

We obtained useful informations from Guy Coignet and Maria Kienzle on the L3
data, from Stefan S\"oldner-Rembold on the OPAL data and we thank them all. One
of us (TTW) is very grateful for hospitality at the CERN Theory Division. This
work was supported in part by the US Department of Energy under Grant
DE-FG02-84ER40158.

\bb{99}
\bi{L3a} L3 Coll., M. Acciari et al., Phys. Lett. {\bf B408} (1997) 450.
\bi{L3b} A. Csilling, L. Fredj and M.N. Kienzle, CERN-EP/L3  Note 2280, June
1998.
\bi{OPAL} F. W\"ackerle, talk given at the XXVII International Symposium
on Multiparticle Dynamics, Frascati, Italy, 8-12 September 1997;
S. S\"oldner-Rembold, talk given at ICHEP'98, Vancouver, Canada, July
22-30, 1998 (hep-ex/9810011).
\bi{CW} H. Cheng and T.T. Wu, Phys. Rev. Lett. {\bf24} (1970) 1456.
\bi{CW69} H. Cheng and T.T. Wu, Phys. Rev. {\bf 182} (1969) 1852.
\bi{CW71} H. Cheng and T.T. Wu, Proceedings of the International Symposium
on Electron and Photon Interactions at High Energies, Cornell Univ.,
Ithaca, N.Y. (1971), N.B. Mistry, M.E. Nordberg and G. Grammer ed.,
p. 147, section 5.
\bi{CWb} H. Cheng and T.T. Wu, Expanding protons : Scattering at high
energies\\
(MIT Press, Cambridge, MA, 1987).
\bi{CWW} H. Cheng, J.K. Walker and T.T. Wu, Phys. Lett. {\bf B44} (1973) 97.
\bi{BSW79} C. Bourrely, J. Soffer and T.T. Wu, Phys. Rev. {\bf D19} (1979)
3249.
\bi{DL} A. Donnachie and P.V. Landshoff, Phys. Lett. {\bf B296} (1992) 227.
\bi{PDG} Review of Particle Physics, The European Physical Journal 
{\bf C3} (1998) 1.
\bi{Eng} R. Engel, Z. Phys. {\bf C66} (1995) 203; R. Engel and J. Ranft,
Phys. Rev. {\bf D54} (1996) 4246.
\bi{Schu} G.A. Schuler and T. Sj\"ostrand, Nuc. Phys. {\bf B407} (1993) 539;
Z. Phys. {\bf C73} (1997) 677.
\bi {Corset} A. Corsetti, R.M. Godbole and G. Pancheri, Phys. Lett.
{\bf B435} (1998) 441.
\bi{Block} M.M. Block, E.M. Gregores, F. Halzen and G. Pancheri,
Phys. Rev. {\bf D58} (1998) 017503.
\bi{BSW94} C. Bourrely, J. Soffer and T.T. Wu, Phys. Lett. {\bf B339} (1994)
322.
\bi{BSW84} C. Bourrely, J. Soffer and T.T. Wu, Nucl. Phys. {\bf B247} (1984)
15.
\bi{Pruss} S.M.Pruss, Proceedings of the VIth Blois Workshop, Blois, 20-24 June
1995, Editions Fronti\`eres 1996, p.3 and references therein.
\bi{H1} S. Aid et al., H1 Coll., Z. Phys. {\bf C69} (1995) 27.
\bi{ZEUS} M. Derrick et al., ZEUS Coll., Z. Phys. {\bf C63} (1994) 391.
\eb
\newpage

\end{document}